\newcommand{\be}{\begin{equation}}
\newcommand{\ee}{\end{equation}}
\newcommand{\bea}{\begin{eqnarray}}
\newcommand{\eea}{\end{eqnarray}}

\documentclass[prb,showpacs]{revtex4}
\usepackage{amssymb}
\usepackage{graphicx}
\usepackage{dcolumn}
\usepackage{bm}
\usepackage{amsmath}
\usepackage{amsfonts}

\begin{document}

\title{Self-consistent approach for calculations of exciton binding energy  in quantum wells}
\author{I. V. Ponomarev}
\email{ilya@physics.qc.edu}
\author{L. I. Deych}
\author{A. A. Lisyansky}

 \affiliation{Department of Physics, Queens College of the City University of New
 York,\\
Flushing, NY 11367}
\date{\today }

\begin{abstract}
We introduce a computationally efficient approach to calculating
the characteristics of excitons in quantum wells.  In this
approach we derive a system of self-consistent equations
describing the motion of an electron-hole pair. The motion in the
growth direction of the quantum well in this approach is separated
from the in-plane motion, but each of them occurs in modified
potentials found self-consistently. The approach is applied to
shallow quantum wells, for which we obtained an analytical
expression for the exciton binding energy and the ground state
eigenfunction. Our results are in excellent agreement with
standard variational calculations, while require reduced
computational effort.
\end{abstract}
 \maketitle

%\preprint{APS/123-QED}

%%%%%%%%%%%%%%%%%%%%%%%%%%%%%%%%%%%%%%%%%%%%%%%%%%%%%%%%%%%%%%%%%%%%%%%

\section{\label{sec:sec1}Introduction}
Excitons play an important role in the band edge optical
properties of low-dimensional semiconductor structures such as
quantum wells, quantum wires, and quantum dots.\cite{Harrison99a}
Quantum confinement of electrons and holes in such structures
results in increased binding energy of excitons, their oscillator
strength, and a life-time. As a result, excitons, for instance, in
quantum wells, are observed even at room temperatures, and play,
therefore, a crucial role in various optoelectronic
applications.\cite{Burstein,Mendez,Nolte} To be able to calculate
effectively and accurately exciton binding energies in the quantum
heterostructures is an important problem, and therefore, a great
deal of attention has been paid to it during several
decades.\cite{Miller81,Bastard82a,Green84,Efros86a,Andreani90a,Gerlach98a,Andreani97a,Kossut97a,
Harrison96a,Laikhtman00a,Ekenberg87,Chang88,Warnock93a,Harrison95a,Voliotis95a,Stahl87,Balslev89a,Merbach98,Castella98}
%Calculations of the exciton binding energy  in various quantum
%well structures with or without external fields have great
%importance for applications.
However, this problem is rather complicated, and while significant
progress has been achieved, new material systems and new type of
applications require more flexible, accurate and effective
methods.

Presently, the best results are usually obtained within the
framework of the variational approach, where a certain form of the
exciton wave function, depending on one or more variational
parameters is being postulated. The exciton energy is then
calculated by minimizing the respective energy functional with
respect to the variational parameters. Unfortunately, even in the
simplest (and, therefore, less accurate) realization of this
approach %for the case of a symmetric single quantum well,
%even in the case of a single
%quantum well
it is  not possible to express a value of the binding energy as a
function of quantum well parameters.  The best one can hope for is
to obtain a set of complicated equations, which relate material
parameters to several variational parameters. The latter are found
numerically, and then used for numerical computation of the
exciton energy. These difficulties result from the fact that
variables in the Hamiltonian describing relative motion of the
electron-hole pair cannot be separated: the presence of the
quantum well potential breaks the translational invariance of the
system, making it impossible to separate in-plane motion of the
electrons and holes from the motion in the direction of
confinement.
 %the center-of-mass and
%relative motions of the electron and the hole. As a result, this
%non-separable Hamiltonian is usually solved numerically from the
%very beginning.

The standard variational approach is based upon a choice of a
specific form of a trial function, which is usually chosen in the
form of a product of three
terms.\cite{Miller81,Bastard82a,Green84,Andreani90a} The first two
are one-particle one-dimensional electron and hole wave functions
for confined motion across the quantum well. The third term
describes the relative motion of an electron and a hole due to
Coulomb interaction. The accuracy of the results depends on the
complexity of the third term of the trial function and the number
of variational parameters. The more parameters, the lower the
binding energy, but, of course, the more  extensive the
calculations. There is ample literature dealing with accurate
variational numerical calculations of exciton binding energy in
quantum
wells.\cite{Miller81,Bastard82a,Green84,Efros86a,Andreani90a,Gerlach98a,Andreani97a,Kossut97a,
Harrison96a,Laikhtman00a,Ekenberg87,Chang88,Warnock93a,Harrison95a,Voliotis95a}.
Most advanced of the calculations include  the effects of Coulomb
screening due to dielectric constant mismatch, as well as
effective mass mismatch at heterojunctions and band degeneracy.
\cite{Andreani90a,Gerlach98a}

Another approach discussed in the
literature\cite{Stahl87,Balslev89a,Merbach98,Castella98} is based
upon an expansion of the electron-hole envelope wave function in
terms of the complete system of eigenfunctions of a one-particle
Hamiltonian describing motion of electrons (holes) in the
respective quantum well confining potentials. The coefficients of
this expansion represent the wave functions of the in-plane
motion. They satisfy an infinite set of differential equations,
which are coupled because of mixing of different electron and hole
sub-bands induced by Coulomb interaction. Such a system can be
solved only numerically after an appropriate truncation of the
basis. Another way is to solve the system in diagonal
approximation and to treat the off-diagonal elements with the help
of  perturbation theory. The improvement of accuracy in this
approach faces difficulties related to the unknown errors due to
basis truncation.

Despite the significant achievement of the current approaches,
they suffer from some principal limitations imposed by their very
nature, and which cannot be, therefore, easily overcome. For
instance, traditional variational approaches are limited by the
need to deal with a variational function of a particular form,
which tremendously restricts the functional space over which the
minimum of the energy is being searched. This problem cannot be
circumvented by an increase in the number of the variational
parameters because of the difficulties solving optimization
problems with three or more parameters. Besides, calculations
presented in most papers are not self-consistent (some limited
attempts to introduce self-consistency, which were made in the
past\cite{Warnock93a,Harrison95a} are discussed below in Section
II). At the same time calculations would become more important for
material systems with wide band-gap materials.
% These may be not a problem for type one $III-V$ quantum wells,
%but one could expect that the self-consistency would become more
%important for material systems with larger exciton effects such as
%wide band-gap materials.
Thus, it is necessary to develop a method
of calculating exciton binding energies, which would be more
flexible and accurate than the existing methods, and which would
allow to treat effects due to electron-hole interaction in a
self-consistent way.

In this work we suggest such an approach, which is based upon
application of the ideas of the self-consistent Hartree method to
the excitons in quantum wells. The idea of this approach is
instead of imposing a particular functional dependence on the
envelope wave function, to  make a more flexible conjecture
regarding the form of this function. Particularly, we present the
total function as a combination of some unknown functions, which
depend on fewer than the total number variables. Applying the
variational principle to this combination we derive a system of
equations describing both the motion of electrons and holes in the
direction of confinement, and the relative two-dimensional
in-plane motion of the exciton. Effective potentials entering
these equations have to be found self-consistently along with the
wave functions.

This approach has a number of advantages compared to the previous
methods. First of all, in its most general statement it must give
better results for the exciton energy because we span a much
larger functional space in the search for the minimum. Second, as
it will be discussed below, this approach automatically gives a
self-consistent description. Third, the approach naturally allows
to incorporate external electric and magnetic fields, stress,
disordered potential acting on electrons and holes in QW because
of inherent inhomogeneities of structure. All these effects, which
modify the single particle part of the Hamiltonian, appear
automatically in self-consistent equations for the variational
functions.
% The idea of this approach is
%to separate variables in the envelope wave function, and to use
%the variational principle in order to derive a system of
% equations describing both the motion of electrons and
%holes in the direction of confinement, and the relative
%two-dimensional in-plane motion of the exciton. Effective
%potentials entering these equations have to be found
%self-consistently along with the wave functions.
We show that for the case of strong confinement in QW the
self-consistent approach with factorized form of the envelope wave
function allows one to achieve a very good agreement with the
results of the variational method for the most elaborate trial
functions used in the literature before.
 The relative motion of the exciton in the in-plane directions
is described in this treatment by the Coulomb potential averaged
with the wave functions of electron's (hole's) motion in the
direction of confinement.  The latter, in turn, is characterized
by an effective confining potential, which is obtained by
combining the initial quantum well potential with the
appropriately averaged Coulomb potential. Unlike the perturbative
method\cite{Stahl87}, our approach takes into account the Coulomb
mixing of the electron and hole sub-bands in a non-perturbative
way, and is expected to give more accurate results even for the
cases when such mixing is important.

We show that the effective exciton potential has two different
regimes  of behavior. For large distances the potential has
three-dimensional Coulomb tails, while at very small distances it
becomes logarithmic as it would be for the true Coulomb potential
of a point charge in two dimensions. A crossover between these two
regimes takes place around distance $r\sim d$, which is the
average electron-hole separation in the quantum well in the
$z$-direction.

Since the main purpose of this paper is to present the new
approach and compare it with the results of calculations carried
out by other methods, we chose a simplified model of a QW
neglecting some of the effects that can be incorporated in the
future. Moreover, we apply our approach to a particular case of a
shallow quantum well, which allows for obtaining some analytical
results, which give important qualitative insight into the
properties of more generic models as well. We use a
$\delta$-functional model to describe shallow quantum wells that
allows us to obtain an explicit form of the effective Coulomb
potential. As a result, we derive a simple analytical formula for
the exciton binding energy that depends only on one variational
parameter. This formula gives results comparable to the best
numerical results obtained by the standard variational approach.
Thus we demonstrate that the method is both efficient and
accurate; it can be applied to any quantum well with an average
size of confinement smaller than the three-dimensional effective
Bohr radius.

The paper is organized as follows. In Sec. \ref{sec:2}, we discuss
the model and derive the general self-consistent equations. In
Sec. \ref{sec:3} we derive an analytical expression and discuss
the different limits of the effective exciton potential. Section
\ref{sec:4} presents the comparison of the
 binding energies results for the self-consistent approach and the
standard variational method. The last section presents the
conclusions of our work. The auxiliary details for the
calculations can be found in three appendices.

\section{The model: 2D exciton in self-consistent field\label{sec:2}}
We assume that both  conduction and valence bands are
non-degenerate, and that they both have an isotropic parabolic
dispersion characterized by the masses $m_e$ and $m_h$ (the heavy
hole mass), respectively.  Throughout the paper we use effective
atomic units (a.u.), which means that all distances are measured
in units of the effective Bohr radius
$a_B=\hbar^2\epsilon/\mu^*e^2$, energies in units of
$\mu^*e^4/\hbar^2\epsilon^2\equiv 2\textrm{ Ry}$, and masses in
units of reduced electron-hole mass $\mu^*$, where
$1/\mu^*=1/m_e^*+1/m_h^*$. In this notation
$m_{e,h}=m_{e,h}^*/\mu^*$, where $m_{e,h}^*$ are effective masses
of an electron and a heavy hole. We assume that both the barrier
and the well have close dielectric constants $\epsilon$
 as well as dispersion laws. Thus, we neglect a dielectric
 constant difference and an effective mass mismatch.
One of the goals of this paper is to compare our method with
existing approaches. Therefore, we have made some of these
simplifications deliberately. Important effects such as
valence-band mixing, non-parabolicity of the conduction band,
dielectric constant and effective mass mismatches can be added to
the model at later stages once the method is fully developed.

After the standard procedure of  excluding the center-of-mass of
the perpendicular motion in the plane of the
layers,\cite{Miller81,Bastard82a} the excitonic Hamiltonian is
given by
\begin{eqnarray}
\hat{H} &=&E_g+H_e+H_h+K_r+V_{reh},\label{H1}\\
H_e &=& -\frac{1}{2m_e}\frac{\partial^2}{\partial z_e^2}
+V_1(z_e),\nonumber\\
H_h &=& -\frac{1}{2m_h}\frac{\partial^2}{\partial z_h^2}
+V_2(z_h),\nonumber\\
K_r &=& -\frac{1}{2}\left[\frac{\partial^2}{\partial
r^2}+\frac{1}{r}\frac{\partial}{\partial r}\right],\nonumber\\
V_{reh} &=& -\frac{1}{\sqrt{r^2+(z_e-z_h)^2}} \nonumber,
\end{eqnarray}
where $E_g$ is a gap energy, $z$ is the growth direction, $r$
measures a relative electron-hole distance in the transverse
direction $r=\sqrt{(x_e-x_h)^2+(y_e-y_h)^2}$, and $V_{1,2}$ are
the quantum well confining potential in $z$ direction for the
electron and the hole, respectively.  We have already assumed that
the ground state must be independent of an angle in the $xy$
plane, and excluded the corresponding term from the kinetic energy
of the relative motion, $K_r$.

A \emph{variational principle} can be used in two different ways
for calculation of approximate solutions for the Schr\"{o}dinger
equation  with the Hamiltonian (\ref{H1}). The first approach is
the \emph{standard variational method}. It is well described in
 the literature.\cite{Miller81,Bastard82a,Green84,Andreani90a,Andreani97a,Kossut97a,
Harrison96a,Gerlach98a,Laikhtman00a,Ekenberg87,Chang88} According
to this method, one needs to start from a variational principle
for the functional $E[\Psi]$:
\begin{equation}\label{funH}
    E\left[\Psi\right]=\int\;\Psi^*\hat{H}\Psi\,dV=\textrm{min}
\end{equation}
with the additional normalization condition
\begin{equation}\label{funHad}
    \int\;|\Psi|^2\,dV =1.
\end{equation}
Then look for an approximate wave function within a class of
functions of predetermined analytical coordinate dependence. These
functions depend on several variational parameters,
$\lambda_1,\lambda_2,\ldots$. Then the total energy
\begin{equation}\label{varmette}
    E=E(\lambda_1,\ldots,\lambda_n),
\end{equation}
and numerical values of variational parameters can be obtained
from minimization conditions
\begin{equation}
    \frac{\partial E(\lambda_1,\ldots,\lambda_n)}{\partial \lambda_i}
    =0,\quad i=1,2,\ldots,n.
\end{equation}
The success of the method depends essentially on the choice of the
trial function. It must be simple enough to lend itself easily to
the calculations, but must vary in a sufficiently large domain for
the energy obtained to be closed to the exact one.

Another way to calculate the approximate solutions of Eq.~
(\ref{H1}) is to utilize the \emph{self-consistent approach}.
 This approach also starts from the \emph{variational principle}, Eqs.~
 (\ref{funH}) and (\ref{funHad}). However, instead of choosing a particular
 coordinate dependence of the trial function, we only assume a particular
functional dependence on different coordinates for the entire wave
function. Namely, we construct an approximate entire wave function
$\Psi(z_e,z_h,r)$ with the help of the unknown functions $\psi_1,
\psi_2,\ldots$, where each function $\psi_k$ depends on a lesser
number of variables than the entire wave function. Considering
variations of these functions independently, from the variational
principle, Eqs.~(\ref{funH}) and (\ref{funHad}), we obtain coupled
integro-differential equations for $\psi_k$.
%
%The self-consistent approach has been employed extensively
%in nuclear, atomic and molecular physics for calculations of many
%particle complex systems. If the trial function of the entire
%system chosen as a linear combination of the products of single
%particle wave functions, it bears the name of self-consistent (or
%mean field) Hartree-Fock approximation.

If localization in the quantum well is strong (the exciton
``$z$-size" is smaller than its Bohr radius), then it is
reasonable to suggest that the exact wave function for the ground
state of Hamiltonian (\ref{H1}) is close to the simple product of
functions of different coordinates
\begin{equation}\label{MFtrialF}
  \Psi_{\textrm{exact}}(r,z_e,z_h)\longrightarrow
  \Psi_{\textrm{trial}}(r,z_e,z_h)=\psi(r)\chi_e(z_e)\chi_h(z_h).
\end{equation}
Assuming normalization of every function in this product, we
substitute function $\Psi$ in Eq.~(\ref{funH}) by the trial
function (\ref{MFtrialF}), vary each function in a product
separately, and obtain the system of coupled integro-differential
equations
\begin{eqnarray}
\left[K_r+\overline{V}_r(r)\right]\psi(r)&=&E_X\psi(r),\label{mfeq1}\\
\left[H_e+\overline{V}_e(z_e)\right]\chi_e(z_e)&=&E_e\chi_e(z_e),\label{mfeq2}\\
\left[H_h+\overline{V}_h(z_h)\right]\chi_h(z_h)&=&E_h\chi_h(z_h),\label{mfeq3}
\end{eqnarray}
where the following notations for effective potentials are
introduced:
\begin{eqnarray}
\overline{V}_r(r)&=&\langle \chi_e\chi_h|V_{reh}|\chi_e\chi_h \rangle,\label{efpotr}\\
\overline{V}_{e,h}(z_{e,h})&=&\langle
\psi\chi_{h,e}|V_{reh}|\psi\chi_{h,e} \rangle. \label{efpotz}
\end{eqnarray}
The angle brackets imply that the integration of the Coulomb
potential with corresponding wave functions is carried out over
two of three independent variables.

 Solving system of equations (\ref{mfeq1})--(\ref{mfeq3}) we obtain the best
approximation for the entire wave function in the form of a
product (\ref{MFtrialF}). The corresponding value of the total
energy is given by Eq. (\ref{funH}) that can be rewritten in the
form
\begin{equation}\label{totalE}
E=\langle \Psi|\hat{H}|\Psi\rangle=E_e+E_h+E_X-
\langle\chi_e|\overline{V}_{e}|\chi_e\rangle-\langle\chi_h|\overline{V}_{h}|\chi_h\rangle.
\end{equation}
The latter expression can be obtained by averaging each of Eqs.~
(\ref{mfeq1})--(\ref{mfeq3}) and adding them together. The
electrostatic term between the electron and the hole is counted
three times in the summations, and so has to be subtracted twice
to give Eq. (\ref{funH}). Thus the total energy is not just the
sum of the exciton binding energy and the electron and the hole
confining energies. The last two terms in Eq. (\ref{totalE})
describe the renormalization of the total energy due to
non-separability of the Hamiltonian.

 In order to solve Eqs. (\ref{mfeq1})--(\ref{mfeq3}) we apply the
method of successive approximations. For strong localization
inside the quantum well, the corrections to the single-particle
energies due to effective Coulomb interaction potentials
$\overline{V}_{e,h}(z_{e,h})$ in Eqs. (\ref{mfeq2}) and
(\ref{mfeq3}) are small. Therefore, we begin by neglecting their
contributions ($\overline{V}^{(0)}_{e,h}$=0) and solve the
equations
\begin{equation}
H_{e,h}\chi_{e,h}^{(0)}(z)=E_{e,h}^{(0)}\chi_{e,h}^{(0)}(z).\label{mfeq20}
\end{equation}
The obtained eigenfunctions are then substituted into Eq.
(\ref{efpotr}) in order to get $\overline{V}^{(0)}_{r}(r)$, a zero
approximation for $\overline{V}_r(r)$. This potential in turn
should be substituted into Eq. (\ref{mfeq1}). The resulting
equation,
\begin{equation}\label{mfeq10}
  \left[K_r+\overline{V}^{(0)}_r(r)\right]\psi^{(0)}(r)=E^{(0)}_X\psi^{(0)}(r),\\
\end{equation}
 describes the formation of a two-dimensional exciton by
an effective electron-hole interaction. The physical meaning of
this effective interaction is  a quantum mechanical average of the
Coulomb potential with confinement wave functions. The ground
state eigenfunction computed from Eq. (\ref{mfeq1}) can then be
substituted into Eq. (\ref{efpotz})  to calculate  a new
approximation $\overline{V}^{(1)}_{e,h}(z_{e,h})$ for the
effective potentials. This process can be continued until the
potentials are self-consistent to a high order of accuracy, i.e.
until the condition
\begin{equation}
\langle\psi|\overline{V}_{r}|\psi\rangle\approx
\langle\chi_e|\overline{V}_{e}|\chi_e\rangle\approx\langle\chi_h|\overline{V}_{h}|\chi_h\rangle
\label{sscond}
\end{equation}
is fulfilled. Eqs.~(\ref{mfeq1})--(\ref{totalE}) with the
condition (\ref{sscond}) represent the complete system of
equations for finding the minimum of the total energy for the
Hamiltonian (\ref{H1}) if the trial function has a particular
functional dependence (\ref{MFtrialF}).

The described procedure has several advantages in comparison with
the standard variational method. First of all, at each step we
solve one-dimensional differential equations.\footnote{We assume
that ground state energy for two-dimensional exciton in the
central field is independent of angles. Therefore, resulting
equation on the radial wave function is one-dimensional.} Second,
even if the resulting Eq. (\ref{mfeq1}) for the exciton in the
effective field cannot be solved analytically, the explicit form
of the effective potential (\ref{efpotr}) gives some additional
understanding of the form of the exciton eigenfunction, and hence
improves the accuracy of calculations.  Finally, the convergence
of the successive iterations itself allows us to estimate to what
degree a given functional dependence of the trial function is
close to the exact wave function.  The energy difference between
the successive approximations shows how far the approximate energy
is from the exact ground state energy. Certainly, we should expect
a slow convergence for very broad and ultra-narrow quantum wells,
where the entire wave function must be close to the wave function
of the three-dimensional exciton. In this case, however, we can
modify our self-consistent theory, rewriting Hamiltonian
(\ref{H1}) in terms of the new independent variables: the
three-dimensional radius $R$ (which is determined by
$R=\sqrt{r^2+z^2}$), the angle $\theta$ (that links $z$ and $R$
coordinates: $z=R\cos\theta$) and the coordinate $Z$ of
center-of-mass in the $z$ direction [$Z=(m_ez_e+m_hz_h)/M$]. Then
we can apply the variational principle for a trial function
$\Psi_{\textrm{trial}}(R,\theta,Z)=\psi(R)f(\theta)g(Z)$, and
obtain corresponding self-consistent equations for functions
$\psi,\;f\;,g$. We do not want to dwell on this issue in this
paper, since our primary interest is mainly focused on experiments
where the average size of the particle localization in the $z$
direction is smaller than the effective three-dimensional Bohr
radius of the exciton. In the next section, we present the
comparison of our self-consistent approach with the results of the
standard variational method for the shallow quantum well. It is
worth noting that the first iteration of our approach might be
considered as an improved version of the standard variational
method with a separable trial function (\ref{MFtrialF}), in which
functions $\chi_{e,h}$ are the electron and the hole
eigenfunctions without the interaction, and ${\psi(r)}$ is chosen
in the form of a $1s$ function for a two-dimensional exciton.

We would like to note that some attempts to treat the Coulomb term
in the Hamiltonian (\ref{H1}) in a self-consistent manner have
been made in the
past.\cite{Efros86a,Warnock93a,Harrison95a,Balslev89a,Merbach98,Castella98}
For example, in Refs.~\onlinecite{Warnock93a,Harrison95a} an
incomplete self-consistent procedure for single-particle wave
functions was performed. Due to the more complicated form of the
trial function
$\Psi_{\textrm{trial}}=\psi(r,z_e-z_h)\chi_e(z_e)\chi_h(z_h)$ the
authors in Refs.~\onlinecite{Warnock93a,Harrison95a}  treated the
first term $\psi(r,z_e-z_h)$ in the product  by the standard
variational method, adjusting variational parameters. Then an
attempt was made to look for self-consistent corrections  to the
electron and hole wave functions, $\chi_{e,h}$, with the help of
equations similar to our Eqs.~(\ref{mfeq2}) and (\ref{mfeq3}). In
principle, it is possible to write the complete system of
self-consistent equations for the trial function in the form
$\psi(r,z_e-z_h)\chi_e(z_e)\chi_h(z_h)$. To do this, one needs to
start again from the variational principle, Eqs.~(\ref{funH}) and
(\ref{funHad}). Then by varying each function separately in the
product one can get the complete system of the
integro-differential equations similar to Eqs.~%
(\ref{mfeq1})--(\ref{mfeq3}). This system and  the final
expression for the total energy will have a more complicated form
due to non-orthogonality of functions in the product for the
entire trial function.

On the other hand, the first iteration of  our method results in
Eq.~(\ref{mfeq10}), which coincides with the zero approximation of
the ``truncated basis" approach,\cite{Balslev89a,Castella98} when
only the ground state confinement eigenfunctions are left. Our
derivation of Eq.~(\ref{mfeq1}) shows, however, that it is more
significant than merely a truncation of all but one term in the
basis. The successive iterations of the Eqs.~%
(\ref{mfeq1})--(\ref{mfeq3}) take into account the Coulomb mixing
of the electron and hole sub-bands in a non-perturbative way, and
give more accurate results even for the cases when such mixing is
important. Numerical calculations confirm that, indeed, Eq.~%
(\ref{mfeq1}) produces results, which are very close to those
obtained by standard variational methods.

\section{Effective potential for exciton in $\delta$-functional shallow well\label{sec:3}}
We define a shallow quantum well as such a well, in which only one
bound state exists for both electrons and holes. In general, the
energy spectrum of the quantum well with height $U$ and finite
length $L$ is $E_n=\pi^2 x_n^2/(2mL^2)$, where $x_n$ are the roots
of the following transcendental equation:
\begin{eqnarray}
  x&=&n-2/\pi\arcsin\left(\frac{\pi}{Lu_0}x\right),\quad n=1,2,\ldots,N\label{transeq}\\
  n-1 &\leq&  x_n \leq n. \nonumber
\end{eqnarray}
Here we introduced a corresponding wave vector $u_0=\sqrt{2mU}$
that characterizes potential height. The number of levels in the
well is given by the condition
\begin{equation}\label{numlevels}
  N=1+\left[\frac{u_0L}{\pi}\right],
\end{equation}
where $[\cdots]$ denotes the integer part of the number. The
condition given by Eq.~(\ref{numlevels}) can also be  interpreted
as a condition of a new level appearing  when the potential grows
in the well. For example, the second level appears when $u_0$ is
equal to the wave vector of the ground state in the infinite well
with the same length $L$: $u_0=\pi/L$.

\begin{figure}[tbp]
\includegraphics{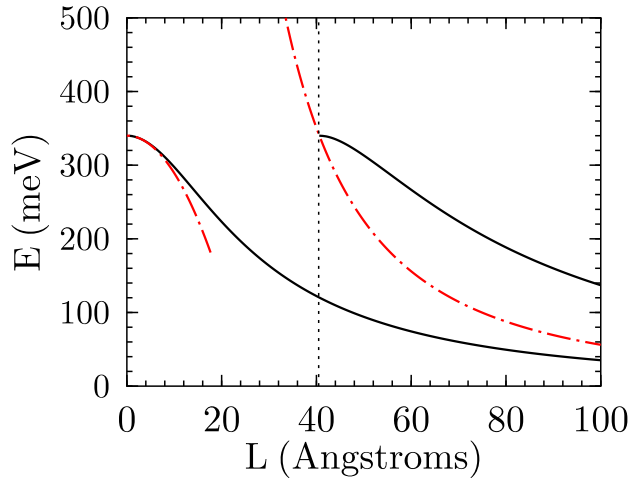}
 \caption{
The energy levels dependence on the well's width. The ground and
the first excited levels (solid lines) are shown for the finite
quantum well. Dot-dashed lines show the ground state level for the
infinite well of the same width and $\delta$-functional potential
(\ref{deltapotential}) with $\alpha=UL$. A vertical dashed line is
the well's width at which the second level in the finite quantum
well appears. It shows the range of applicability of the
$\delta$-functional potential with the effective strength
$\alpha=UL_{\textrm{eff}}$. Parameters are taken for a conduction
band electron in  an Al$_{0.3}$Ga$_{0.7}$As/GaAs quantum well (see
below in a text).} \label{fig1}
\end{figure}

 A transcendental form of Eq. (\ref{transeq}) as well as a piecewise
character of  the eigenfunctions present additional obstacles for
further calculations of the exciton binding energies. Therefore,
different approximations of the finite quantum well
 are often used. For a wide quantum well with several energy
 levels inside, the model of an infinite quantum well with a slightly
 larger effective length $L_{\textrm{eff}}=L/x_1$ is an appropriate one.\cite{Miller85a}
 It gives the same ground energy and correct wave function
 behavior. However, for shallow quantum wells with one level
 inside, the use of this model is not justified. Indeed, for an infinite
 quantum well the ground state energy  grows with the decrease of
the  well's width, while the ground state energy in the finite
well has a different dependence, and tends
 to the finite limit $U_0$ when the width tends to zero:
 $E_1\longrightarrow U-U^2L^2m/2$.
 Narrow quantum wells have a different analytical limit of the $\delta$-functional
 potential,\cite{Andreani97a}
\begin{equation}\label{deltapotential}
  V(z)=U-\alpha\delta(z),
\end{equation}
where $\alpha$ is a $\delta$-potential strength. If we define this
parameter as
\begin{equation}\label{alphaeffective}
  \alpha=UL_{\textrm{eff}}=\sqrt{\frac{2U}{m}-\frac{\pi^2x_1^2}{m^2L^2}}
\end{equation}
where $L_{\textrm{eff}}$ is
 chosen to match the ground state energy of the finite well problem,
then the well-width range of applicability of this approximation
is extended up to the moment of the appearance of the second level
in the finite quantum well. For typical parameters in AlGaAs/GaAs
structures it corresponds to a well size
 $L\approx 40\AA$. Obviously, $L_{\textrm{eff}}\rightarrow L$ when $L$
tends to zero. Figure~\ref{fig1} shows typical energy dependence
on well's width for the electron in the AlGaAs/GaAs quantum well
for the finite quantum well and its approximations. Comparing the
behavior of curves for the ground state  of the finite width well
and its $\delta$-functional approximation with strength
$\alpha=UL$,  one can see that $\delta$-functional curve stays
always on the left. It means that the effective length parameter,
determined by Eq. (\ref{alphaeffective}) should be smaller than
the actual well width, which is opposite to the case of effective
infinite quantum well width. In some sense, the model of
$\delta$-functional QW is complimentary to the model of effective
infinite quantum well\cite{Miller85a} (EIQW), which is used to
approximate finite QW with large widths (and/or barrier heights),
when the number of levels in a well is large. Indeed, the more
discrete levels exist in the QW the better the EIQW model works,
but it fails gives a wrong eigenstate dependence on $L$, when the
well has only one level. On the other hand the $\delta$-functional
QW is not applicable for quantum wells with more than one level.
The delta-functional approximation is applicable both to very
narrow QW and to wells with a small band-gap
offset,\cite{Andreani97a} i.e. when the well width and/or the band
offsets are very small so that the carrier wave functions are
mostly in the barrier region.
%%%%%%%%%%%%%%%%%%%%%%%%%%%%%%%%%%%%%%%%%%%%%%%%%%%%%%%%%%%%%%%%%%%
%%%%%%%%%%%%%%%%%%%%%%%%%%%%%%%%%%%%%%%%%%%%%%
For the $\delta$-functional potential it is more convenient to
count energies from the barrier band edge rather than from the
bottom of the well. In terms of the total Hamiltonian (\ref{H1})
it means that the energy band gap constant is the barrier's energy
band gap: $E_g^{\textrm{bar}}=E_g^{\textrm{well}}+U_e+U_h$. The
energy and wave function of a single localized state are
well-known:
\begin{equation}
E_{e,h}=-\frac{\kappa_{e,h}^2}{2m_{e,h}},\qquad \chi_{e,h}(z)
=\sqrt{\kappa_{e,h}}\exp(-\kappa_{e,h}|z|)\label{snglwellwf}.
\end{equation}
Parameters $\kappa_{e,h}=\alpha_{e,h}m_{e,h}$ determine the
localization of  wave functions of an electron and a hole,
respectively. It is worth noting that even for a very shallow
quantum well, the localization length ($\sim 1/\kappa$) might be
much less than the effective Bohr radius. In this case, we can
expect a quasi-two dimensional behavior for the exciton,
justifying the approximation for the mean field function in  form
(\ref{MFtrialF}). In the case of the AlGaAs/GaAs quantum wells,
the electron (hole) localization length is smaller than Bohr's
radius up to $L\approx 5\AA$.

\begin{figure}[tbp]
\includegraphics{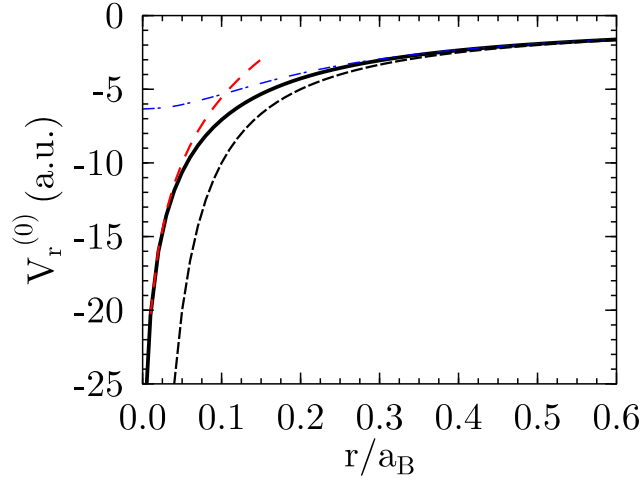}
\caption{An effective self-consistent potential
$\overline{V}^{(0)}_r$ profile for a two-dimensional exciton. The
solid thick line represents Eq. (\ref{Vr0}). The dashed-dotted
line is the approximation $(r^2+d^2)^{-1/2}$, the dashed line is
the logarithmic regime Eq. (\ref{Veffsm}) for small $r$, the
dotted line represents $1/r$ behavior. All data are for
$d/a_B=0.16$ corresponding to  $L=20\AA$ finite quantum well in
AlGaAs/GaAs materials.} \label{fig2}
\end{figure}

 With the help of the wave functions (\ref{snglwellwf}), it is possible
 to obtain the analytical expression for the effective exciton
 potential $\overline{V}^{(0)}_r(r)$. The details of these calculations
 are given in  Appendix~\ref{sec:Ap1}. The result is
\begin{equation}\label{Vr0}
\overline{V}^{(0)}_r(r;\kappa_e,\kappa_h)=-\frac{2\kappa_e\kappa_h}{\kappa_h^2-\kappa_e^2}
\left[\kappa_h T(\kappa_e r)- \kappa_e T(\kappa_h r)\right],
\end{equation}
where the function $T(\kappa r)$ is a combination of zeroth-order
Struve and Neuman functions:\cite{Abramowitz}
\begin{equation}\label{Vr0b}
T(\kappa r)=\frac{\pi}{2}\left[\textbf{H}_0(2\kappa
r)-\textrm{Y}_0(2\kappa r) \right].
\end{equation}
 The behavior of the potential (\ref{Vr0}) is
shown in Figure~\ref{fig2}.

%%%%%%%%%%%%%%%%%%%%%%%%%%%%%%%%%%%%%%%%%%%%%%%%%%%%%%%%%%%%%%%%%%%
%%%%%%%%%%%%%%%%%%%%%%%%%%%%%%%%%%%%%%%%%%%%%%
The behavior of the potential (\ref{Vr0}) has two regimes that are
determined by the parameter
\begin{equation}\label{dkappa}
  d=\sqrt{\frac{1}{2}\left(\frac{1}{\kappa_e^2}+\frac{1}{\kappa_h^2}\right)}.
\end{equation}
This parameter has the meaning of an average electron-hole
separation in the $z$ direction. For large distances,  potential
(\ref{Vr0}) has asymptotic behavior
\begin{equation}\label{Veffasym}
V_{\textrm{asym}}(r)\approx -\frac{1}{\sqrt{r^2+d^2}},\qquad
r\gtrsim d.
\end{equation}
At small distances, the attraction becomes stronger. It has
logarithmic behavior:\footnote{For $\kappa_e\neq \kappa_h$ the
series expansion of Eq.~(\ref{Veffsm}) gives
$d=(\kappa_e^{-1}+\kappa_h^{-2})/2$, which is different from the
definition (\ref{dkappa}). However, the discrepancy between these
two definitions is  negligible for the whole range of
$\kappa_{e,h}$ under the interest.}
\begin{equation}\label{Veffsm}
V_{\textrm{sm}}(r)\approx \frac{1}{d}\left[\ln(r/d)-1+\gamma
\right],\qquad r\lesssim d,
\end{equation}
where $\gamma=0.5772$.%

Thus, the effective electron-hole interaction for the exciton in
the quantum well starts from the true logarithmic Coulomb
potential of a point charge in two dimensions that smoothly
transforms at distances $r\sim d$ to the screening potential
(\ref{Veffasym}) with three-dimensional Coulomb tails. For the
strong confinement $d\ll 1$ we can approximate
$V_{\textrm{eff}}(r)$ by Eq. (\ref{Veffasym}) for all distances
and take into account the logarithmic part on the next step as a
perturbation.

It is interesting to note that potential (\ref{Veffasym}) can be
obtained without the self-consistent procedure from the following
simple intuitive consideration. At the first step, lets us neglect
the electron-hole interaction in Hamiltonian (\ref{H1}). Then, we
can solve the one-dimensional one-particle Schr\"{o}dinger
equations in the quantum well, and find the average square of the
distance between the electron and the hole as
\begin{equation}\label{avdist}
  d^2=\langle
  \chi_e(z_e)\chi_h(z_h)|(z_e-z_h)^2|\chi_e(z_e)\chi_h(z_h)\rangle.
\end{equation}
This yields the same result as Eq.~(\ref{dkappa}). The next step
in the approximation of the Hamiltonian (\ref{H1}) is to include
the Coulomb attraction term, where $(z_e-z_h)^2$ is substituted by
its average value $\langle \ (z_e-z_h)^2\ \rangle$.

\section{Numerical results: comparison with standard variational approach\label{sec:4}}
The Schr\"{o}dinger equation for a radial wave function in the
central field (\ref{Vr0}) does not have an analytical solution. To
obtain an approximate analytical expression for the exciton
binding energy we will use the following procedure.

First of all, let us stress again that despite the fact that we
consider a shallow quantum well with one single particle
eigenvalue inside, the strong confinement persists up to very
small widths. The parameter $d$ is a good indicator of such
confinement. For example, in Table~\ref{tab:table1} the data are
presented for Al$_{0.3}$Ga$_{0.7}$As-GaAs materials. We can see
that for the well's width of $10\AA$ this parameter is about one
quarter of the three-dimensional  Bohr radius and even smaller for
larger quantum wells. For the case of strong confinement, $d\ll
1$, the effective potential (\ref{Vr0}) can be represented by
Eq.~(\ref{Veffasym}) almost everywhere. Therefore, at the first
step, it is reasonable to substitute the potential (\ref{Vr0}) by
its asymptotic form (\ref{Veffasym}) for all distances.

 Appendix~\ref{sec:Ap3} yields
the details of numerical calculations and analytical limits for
the ground state in the potential. Although the Schr\"{o}dinger
equation  with the potential (\ref{Veffasym}) also does not have
an analytical solution, we discovered that the ground state
energy, obtained  by the variational method for the single
parameter trial function
\begin{equation}\label{trpsi_scheq}
\varphi_{trial}=
\frac{2\exp(d/\lambda)}{\sqrt{\lambda(\lambda+2d)}}
\exp(-\sqrt{r^2+d^2}/\lambda),
\end{equation}
coincides with the exact one with excellent accuracy. To check
this, we performed a precise numerical integration of the
Schr\"{o}dinger equation based on Pruefer transformation and a
shooting method. The difference in the ground state energies for
the whole range of the parameter $d$ was less than $ 0.01\%$, or
$\sim 10^{-3}$ meV! Such an  agreement can be explained by the
fact that trial function (\ref{trpsi_scheq}) has a correct
analytical behavior for both  small and large distances, $r$.

\begin{figure}[tbp]
\includegraphics{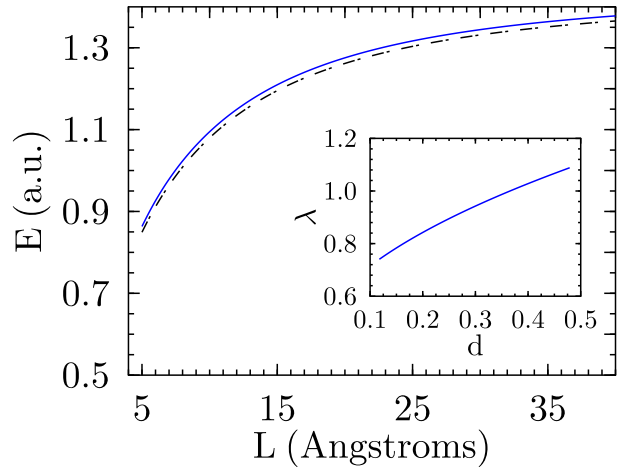}
 \caption{The exciton binding energies in the shallow quantum well for
 different well widths $L$. The solid line represents the two-dimensional
 exciton ground state, Eqs.~(\ref{excenergy_sc0}) and (\ref{excenergy_sc1}) in the
 effective potential. The
 dotted-dashed line is the exciton binding energy obtained with the
 help of the variational method with the trial function (\ref{tf01a}).
 The insert shows the dependence of the parameter $\lambda$ on the average
 electron-hole separation $d$ in Eq.~ (\ref{excenergy_sc0}).
 Both parameters are expressed in Bohr radius units and are presented for
  the same range of quantum well widths.} \label{fig2b}
 \end{figure}

The expression for the ground state energy obtained for the trial
function (\ref{trpsi_scheq}) is given by
\begin{equation}\label{excenergy_sc0}
E_X^{(0)}(\lambda)=-\frac{2}{\lambda}\frac{1}{1+2d/\lambda}+\frac{1}{\lambda^2}
\left[1-\frac{(2d/\lambda)^2E_1(1,2d/\lambda)\exp(2d/\lambda)}{1+2d/\lambda}\right],
\end{equation}
where $\textrm{E}_1(x)$ is the exponential
integral.\cite{Abramowitz} The variational parameter $\lambda$
changes from $1.1$ to $0.74$ when the average electron-hole
distance $d$ varies from 0.11 to 0.48. The latter corresponds to
the quantum well widths range from $40\AA$ to $5\AA$ for the
AlGaAs/GaAs structures. The behavior of the parameter $\lambda$ as
a function of $d$ is shown in the insert of Fig.~\ref{fig2b}. At
small $d$ it has the following form:
\begin{equation}\label{lambdavsd}
\lambda\approx \frac{1}{2}+4d+\frac{d^2}{\lambda^2}
\left[4\ln\left(\frac{2d}{\lambda}\right)-12\lambda +4\gamma+1
\right].
\end{equation}
At small distances the effective potential (\ref{Vr0}) differs
from Eq.~(\ref{Veffasym}). The correction to the energy due to
this difference  can be taken into account with the help of
perturbation theory:
\begin{eqnarray}
E_X^{(1)}&=&\langle \varphi_{trial}|\Delta
V|\varphi_{trial}\rangle \approx
\int_0^{d}\varphi_{trial}^2(r)\left[\frac{1}{d}\ln(r/d)\right]r\,dr\label{excenergy_sc1}\\
&=&-\frac{2d}{\lambda^2}\left[\frac{1}{2}-0.557\frac{2d}{\lambda}+0.563
\left(\frac{2d}{\lambda}\right)^2+\cdots\right].\label{excenergy_sc1a}
\end{eqnarray}
The last column of the Table~\ref{tab:table1} represents the final
sum  $E_X^{sc}=E_X^{(0)}+E_X^{(1)}$ for the exciton binding energy
obtained by the self-consistent approach.
\begin{table}[ptb]
\caption{Effective parameters of single quantum well potentials,
electron, hole and exciton binding energies for different quantum
well widths, $L$. Single particle electron and hole energies,
$E_{e,h}^{(0)}$, are the initial step in the self-consistent field
iterations [see Eq.~(\ref{mfeq20})] when effective Coulomb terms
$\overline{V}_{e,h}$ are omitted. These energies determine the
strength  parameters, $L_{\textrm{eff}}^{e,h}$, for  corresponding
$\delta$-potentials and the average distance, $d$, between an
electron and a hole [Eq.~(\ref{dkappa})]. Energies
$E_X^{(1),(2),(3)}$ are the exciton binding energies obtained by
variational method with the trial functions given by Eqs.~%
(\ref{tf01a}),(\ref{tf02a}), and (\ref{tf03a}), respectively. The
last
column shows the energy $E_X^{sc}=E_X^{(0)}+E_X^{(1)}$ [see Eqs.~%
(\ref{excenergy_sc0}),(\ref{excenergy_sc1})] obtained by the first
iteration of the self-consistent approach. The calculations are
based on the following physical constants:\cite{Miller85a}
$m_e^*=0.067m_0$, $m_h^*=0.45m_0$, $U_e=340$ meV, $U_h=70$ meV,
$\epsilon=13.8$. For these parameters the effective Bohr radius is
$a_B=125\AA$ and the energy atomic unit is equal to 8.33 meV.
 }%
\label{tab:table1}%
\begin{ruledtabular}
\begin{tabular}{ccccccccccc}
$L\ (\AA) $ & $L_{\textrm{eff}}^{e}\ (\AA)$ &
$L_{\textrm{eff}}^{h}\ (\AA)$ & $d (a.u.)$ & $-E_e^{(0)}$ (a.u.) &
$-E_h^{(0)}$ (a.u.)& $-E_X^{(1)}$ (a.u.) &$-E_X^{(2)}$ (a.u.)
&$-E_X^{(3)}$ (a.u.) &$-E_X^{\textrm{sc}}$ (a.u.)

\\ \hline
10 & 9.15  & 8.89  & 0.257 & 5.11 & 1.37 & 1.0799 &1.0975  & 1.1041 &1.0948\\
20 & 15.16 & 14.12 & 0.158 & 14.02 & 3.46& 1.2617 &1.2779  & 1.2843 &1.2754\\
30 & 18.63 & 16.88 & 0.130 & 21.18 & 4.95& 1.3314 &1.3467  & 1.3527 &1.3443\\
40 & 20.70 & 18.43 & 0.118 & 26.12 & 5.90& 1.3655 &1.3802  & 1.3855 &1.3779\\
\end{tabular}
\end{ruledtabular}
\end{table}
To check the accuracy of our method we compared the results of the
self-consistent approach with the results of the standard
variational method for three different trial functions. These
trial functions  have the following forms:
\begin{eqnarray}
\psi^{(1)}(z_e,z_h,r;\lambda)&=&\sqrt{\kappa_{e}\kappa_{h}}
\exp(-\kappa_{e}|z_{e}|-\kappa_{h}|z_{h}|) \exp(-r/\lambda),\label{tf01a}\\
\psi^{(2)}(z_e,z_h,r;\lambda,\beta)&=&\sqrt{\kappa_{e}\kappa_{h}}
\exp(-\kappa_{e}|z_{e}|-\kappa_{h}|z_{h}|)\exp(-\sqrt{r^2+\beta^2}/\lambda),\label{tf02a}\\
\psi^{(3)}(z_e,z_h,r;\lambda,\beta)&=&\sqrt{\kappa_{e}\kappa_{h}}
\exp(-\kappa_{e}|z_{e}|-\kappa_{h}|z_{h}|)\exp(-\sqrt{r^2+\beta^2(z_e-z_h)^2}/\lambda),
\label{tf03a}
\end{eqnarray}
The first two functions are separable, while the third one is
non-separable. The first wave function has one variational
parameter $\lambda$, and two others have two variational
parameters $\lambda$ and $\beta$.

The details of variational calculations are given in
Appendix~\ref{sec:Ap2}. Our results and their comparison with data
obtained by the standard variational procedure are presented in
Table~\ref{tab:table1}.

We can see that the single parameter separable trial function,
$\psi^{(1)}(z_e,z_h,r;\lambda)$, gives higher binding energy with
maximal relative difference of $1.3\%$. The non-separable trial
functions have slightly lower binding energies than the first
iteration of the self-consistent approach. The maximum relative
difference between the self-consistent binding energy $E_X^{sc}$
and the binding energy $E_X^{(2)}$ is $0.24\%$ (for AlGaAs/GaAs
structures it corresponds to a difference of $0.02$ meV). For
$E_X^{(3)}$ it is $0.84\%$ ($0.07$ meV).
%
%It is worth noting that%
These results demonstrate that even the first iteration of our
approach gives an excellent agreement with the variational
results. Subsequent iterations further decrease the ground state
energy. Preliminary research on the convergence of the
self-consistent approach shows that the next iteration gives the
value of the binding energy lower than the variational
calculations with trial functions (\ref{tf02a}) and (\ref{tf03a}).
These results will be published elsewhere.
%
%We would also like stress that the results for already the first
%iteration coincide with the results of variational results with a
%very good accuracy.
However, in the particular case considered here the difference by
$0.07$ meV between more complicated variational approach and the
results presented here is already much smaller than the accuracy
of, for instance,  optical absorption
experiments\cite{Voliotis95a}, which is of the order of $2$ meV.
From this point of view the discrepancy between the two methods is
negligible. At the same time, our approach is about $100$ times
faster than the variational method even with the simple enough
trial function, Eq.~(\ref{tf02a}). It also helps to avoid a
numerically difficult task of finding minima of several
non-polynomial functions. The additional physical information
about the effective potential allows one to find one of the
variational parameters of function (\ref{tf02a}) without
minimization.
%principles.In our approach we are able to reduce
  %In many cases such accuracy could be
%excessive. Indeed, theoretical
%estimates that includes
%the mass or dielectric constants mismatch, give corresponding
%corrections to the binding energy of the order of 0.5-5 meV. On
%the other hand, an experiment on the optical absorption
%itself\cite{Voliotis95a} gives an error of the order of 2 meV. On
%this background, results obtained either by the variational method
%or our calculations could be considered as identical.
%The only preference for choosing the method is determined
%by simplicity and the efficiency. To compare the
%computational efforts, we present the following example. All our
%calculations were made with the help of Maple software on a
%standard personal computer with Pentium III processor. It took
%about a 1 minute to find the minimum for a simple enough trial
%function Eq. (\ref{tf02}). It took less than 1 second to find the
%minimum of Eq. (\ref{excenergy_sc0}), to sum it with Eq.
%(\ref{excenergy_sc1}), and to get the result in the last column of
%Table~\ref{tab:table1}.
%
%Thus, the comparison of the results shows that the self-consistent
%exciton binding energy obtained from Eqs.~(\ref{excenergy_sc0})
%and (\ref{excenergy_sc1a}) matches the binding energies
%numerically calculated from the standard variational method with
%excellent accuracy.  Moreover, its relative simplicity and
%efficiency give a hope that its value will increase in
%applications for more sophisticated physical problems.
%

%%%%%%%%%%%%%%%%%%%%%%%%%%%%%%%%%%%%%%%%%%%%%%%%%%%%%%%%%%%%%%%%%%%
\section{Conclusions\label{sec:conclusions}}
We introduce a self-consistent approach for calculations of the
exciton binding energy in a quantum well. For the case of strong
confinement, the self-consistent Hamiltonian is separable and
consists of three parts: one-dimensional Hamiltonians for electron
and hole confined motions across the quantum well, and the
Hamiltonian, which describes the motion of a two-dimensional
exciton in the effective central field potential. This effective
potential is a result of averaging over $z$ coordinates of the
Coulomb interaction and the quantum well potential. As a function
of distance the effective potential has two different regimes of
behavior which are determined by the average distance between
electron and hole inside the quantum well, $d$. For small
distances, $r\lesssim d$, the effective potential has a
logarithmic form of a Coulomb potential of a point charge in two
dimensions. At a distance $r \sim d$ this behavior crosses over to
the three-dimensional Coulomb screened potential,
$-1/\sqrt{r^2+d^2}$. For the case of the shallow quantum well,
analytical formulae, Eqs.~(\ref{excenergy_sc0}) and
(\ref{excenergy_sc1}), for the exciton binding energy are
obtained. Even though the use of these formulae requires
computational time which is by orders of magnitude smaller than
that of standard variational calculations, the results obtained by
both methods are in an excellent agreement. The differences
between the exciton binding energies in two approaches are
generally smaller than $1\%$. For AlGaAs/GaAs structures it
corresponds to differences smaller than $0.1$ meV, i.e. smaller
than corrections\cite{Andreani90a} due to non-parabolicity of the
bulk conduction band or dielectric constant and effective mass
mismatches. One can expect that the developed method can lay the
foundation for  models that incorporate these important effects as
well as take into account an influence of the external electric
field.

All results discussed above are obtained for the first successive
iteration. Obviously, the next iterations will lower the ground
state energy even more. For example, the next step is the
substitution of the wave function (\ref{trpsi_scheq}) into Eq.~%
(\ref{efpotz}), which is approximately equivalent to the
appearance of  additional oscillatory potentials in
Eqs.~(\ref{mfeq2}) and (\ref{mfeq3}):
\begin{equation}\label{correction2}
  \overline{V}_{e,h}(z_{e,h})\approx
  -\overline{(r^2+d^2)^{-1/2}}+\frac{\overline{(r^2+d^2)^{-3/2}}}{2}\left(z_{e,h}^2-\overline{z^{2}_{e,h}}\right).
\end{equation}
These potentials will slightly change the single-particle energies
and will localize the tails of the eigenfunctions of the electron
and the hole ground states. The latter will be manifested in a
decrease of the average electron-hole distance $d$, and,
therefore, will lower the exciton binding energy in the next
successive iteration of  Eq.~(\ref{mfeq1}).

%We would like to note that the developed method is flexible and
%useful for real existing needs in excitonic systems or
%exciton-based applications.
In derivation of our results we used three different
approximations.
%Below, we  would like to distinct these
%consecutive approximations and a range of their applicability for
%the Schr\"{o}dinger equation:
They are: ({\it i}) the self-consistent approach itself, ({\it
ii}) the use of the factorized form of the wave function, %made
%within self-consistent approach,
({\it iii}) and the use of the delta-functional potential for a
shallow quantum well. The self-consistent approach is broader than
the standard variational method since we do not have to specify a
particular functional dependence for a trial function.
%put much
%less restrictions on the trial function.
Instead,
%of the particular functional dependence for the trial
%function,
we just suggest that the trial function consists of some
combination of unknown functions. The factorized form is the
simplest form for such a combination but it is not required by our
method.
%for a particular model we can construct a particular combination.
The method can be applied to other physical models where different
types of trial functions would be more natural. For example, for a
wide double quantum well, the better choice of the trial function
of the ground state is a superposition of two factorized
single-well functions. It will result in a system of coupled
equations similar (but more complicated) to
Eqs.~(\ref{mfeq1})--(\ref{mfeq3}). This approach can also be
straightforwardly expanded to include other effects, which were
neglected in this paper. For instance, in order to take into
account the dielectric
mismatch,\cite{Andreani90a,Laikhtman00a,Gerlach98a} we would need
to correct the expression for the effective potentials,
Eq.~(\ref{efpotr}) and Eq.~(\ref{efpotz}), including effects of
image charges into the respective integrals. The valence band
degeneracy and anisotropy can be included by introducing a four
component trial function, for which the self-consistent equations
will have the similar form as Eqs.~(\ref{mfeq1})--(\ref{mfeq3}),
but they should be understood as matrix equations.

%But if we will be able to
%calculate the effective potentials for this system again, it might
%bring additional simplifications. We had also mentioned at the end
%of section~\ref{sec:2} that for cases closed to three-dimensional
%excitons another factorizations is useful.

Turning to the particular factorization used in our problem, it is
obvious that the more strongly the exciton is localized in the
$z$-direction (size of wave function in $z$-direction compared to
the three-dimensional Bohr radius) the better our approximation
works. If we consider the case of Al$_{0.3}$Ga$_{0.7}$As-GaAs
quantum well, it means that our method will work for any quantum
well with a width less that $100\ \AA$, but the best convergence
will happen somewhere around $30-50\ \AA$. Correspondingly, the
factorized form of the trial function for the self-consistent
method is applicable for any structure (e.g., asymmetric quantum
well or quantum well in electric field), if a  wave function
localized in $z$-direction has such an extension. In a similar
matter the self-consistent approach can be applied to the lower
dimension systems such as quantum wires and quantum dots.
Moreover, preliminary consideration showed that the problem of
divergency of the exciton ground state in a one-dimensional
Coulomb potential, which arises in other approaches to quantum
wires, in this method does not appear at all.

The $\delta$-functional potential is a good approximation if a
one-dimensional quantum well has only one level (is shallow). For
a typical case of Al$_{0.3}$Ga$_{0.7}$As-GaAs quantum well, it
gives an applicability range of $L<40\ \AA$. The approximation of
the $\delta$-functional potential gives simple single-particle
wave functions that significantly simplify calculations of the
effective potentials in Eqs.~(\ref{efpotr}) and (\ref{efpotz}). If
asymmetric quantum well or double quantum wells, or quantum well
in electric field are shallow, the delta-functional approach will
be applicable to such models.
% The delta-functional
%potential is also applicable to various types of shallow wells,
%e.g., an asymmetric quantum well, two double quantum wells, or a
%quantum well in electric field.
An asymmetric quantum well can be modelled by a quantum barrier
and the $\delta$-functional potential ($\delta(z) \rightarrow
\delta(z)[1+A\theta(z)]$, where $\theta$ is the step function). An
advantage of using the $\delta$-functional potential is especially
clear in the case of the quantum confined Stark effect for a
shallow quantum well, where it allows one to obtain additional
results related to the field-induced resonance widths for a
single-particle as well as for the exciton quasi-bound states.
These results
%for a shallow quantum well in the external electric field
will be published elsewhere.
%The approach is also applicable for calculations of excitonic effects
%in lower dimensions like quantum wires and dots, where the
%effective potential, calculated in the similar way, introduces a
%natural cut off for the ground energy renormalization.

\begin{acknowledgments}
We are grateful to S. Schwarz for reading and commenting on the
manuscript. The work is supported by AFOSR grant F49620-02-1-0305
and PSC-CUNY grants.
\end{acknowledgments}

\appendix
\section{\label{sec:Ap1} Effective potential for 2D exciton}
The effective field for quasi 2D exciton is given by the integral
\be\label{Veff01} V_{\textrm{eff}}(r)=-\int_{-\infty}^{\infty}dz_e
\int_{-\infty}^{\infty}dz_h  \frac{\left|\chi_e(z_e)\right|^2
\left|\chi_h(z_h)\right|^2}{\sqrt{r^2+(z_e-z_h)^2}}. \ee For a
shallow quantum well approximated by the $\delta$-function
potential it gives
\begin{equation}\label{Veff02}
  V_{\textrm{eff}}(r)=
 -\int_{0}^{\infty}dz_1 \int_{0}^{\infty}dz_2
 e^{-2\kappa_1 z_1}e^{-2\kappa_2 z_2}
 \left[\frac{2\kappa_1 \kappa_2}{\sqrt{r^2+(z_1-z_2)^2}}+\frac{2\kappa_1 \kappa_2}{\sqrt{r^2+(z_1+z_2)^2}}
 \right] =V_1+V_2.
\end{equation}
 After making the coordinate transformation $\xi=z_1-z_2,\ \eta=z_1+z_2$ and
 taking into account that
\begin{equation}\label{intlimits}
\int_{0}^{\infty}dz_1 \int_{0}^{\infty}dz_2 f(\xi,\eta)=
 \frac{1}{2}\int_{0}^{\infty}d\eta \int_{-\eta}^{\eta}d\xi f(\xi,\eta)
 =\frac{1}{2}\int_{0}^{\infty}d\xi \int_{\xi}^{\infty}d\eta\left[
 f(-\xi,\eta)+f(\xi,\eta)\right],
\end{equation}
the first integration in these two interated integrals becomes
trivial and the second integration can be expressed through the
function
\begin{equation}\label{TE}
T(\kappa r)=\int_{0}^{\infty} \frac{\exp(-2\kappa
t)dt}{\sqrt{r^2+t^2}}=\frac{\pi}{2}\left[\textbf{H}_0(2\kappa
r)-\textrm{Y}_0(2\kappa r) \right],
\end{equation}
where $\textbf{H}_0$ is the zeroth-order Struve function and
$\textrm{Y}_0$ is the zero-order Neumann or Bessel function of the
second kind.\cite{Abramowitz} Then potentials $V_1$ and $V_2$ can
be expressed as
\begin{equation}\label{V1V2}
V_{1,2}=-\frac{\kappa_1\kappa_2}{\kappa_1 \pm \kappa_2}
 \left[T(\kappa_2 r)\pm T(\kappa_1 r)\right],
\end{equation}
and the final result yields Eq.~(\ref{Vr0})
\begin{equation}\label{Veff03}
V_{\textrm{eff}}(r;\kappa_1,\kappa_2)=-\frac{2\kappa_1\kappa_2}{\kappa_2^2-\kappa_1^2}
\left[\kappa_2T(\kappa_1 r)- \kappa_1 T(\kappa_2 r)\right].
\end{equation}
In the case when $\kappa_1=\kappa_2\equiv\kappa$ this expression
is reduced to
\begin{equation}\label{Veff04}
V_{\textrm{eff}}(r;\kappa)=-\kappa \left[T(\kappa r)-\kappa
T^{\prime}(\kappa r) \right],
\end{equation}
where
\begin{equation}\label{Tprime}
T^{\prime}(\kappa r)=\frac{\partial T}{\partial \kappa}
 =2r\left[\frac{\pi}{2}\textrm{Y}_1(2\kappa r) -\frac{\pi}{2}\textbf{H}_1(2\kappa
r)+1 \right].
\end{equation}
\section{\label{sec:Ap3}Eigenvalues and eigenfunctions in $2D$ central potentials}

\begin{figure}\label{fig3}
\includegraphics{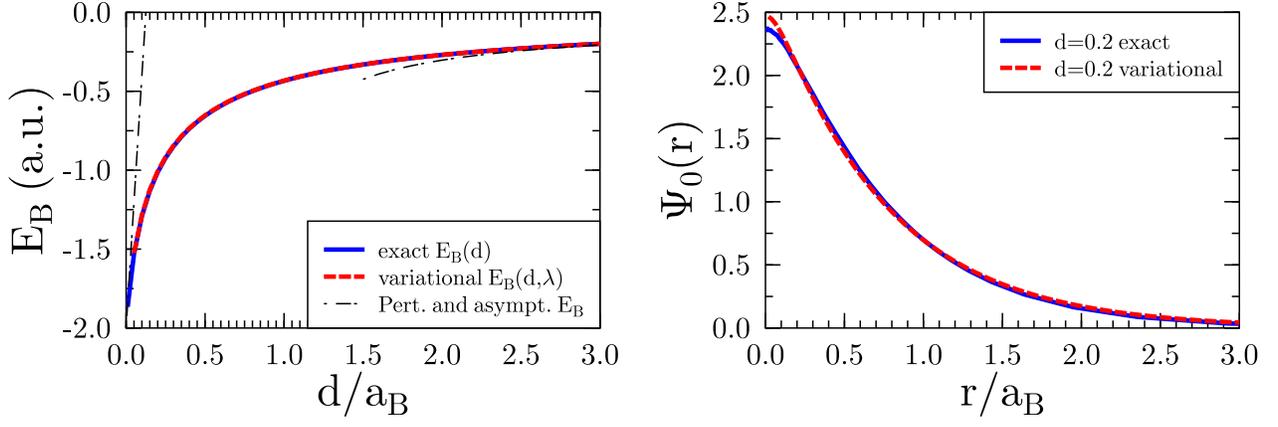}
%\centerline{\epsffile{fig5.eps}}
\caption{ (a) The ground state energy for the potential
$-(r^2+d^2)^{-1/2}$, Eq.~(\ref{Pon}), as functions of the
parameter $d$. Solid line, representing numerical results, and
dashed line, obtained by the use of variational formula
(\ref{enerexpect}), are practically indistinguishable. The
dot-dashed lines correspond to perturbation theory for small $d$
and an asymptotic oscillatory energy $-d^{-1}+d^{-3/2}$ for $d\gg
1$. (b) The ground state wave function for $d=0.2$. The numerical
results are shown by the solid line. The  dashed line corresponds
to Eq.~(\ref{varmethtf}) for normalized trial function.}
\end{figure}

The Schr\"{o}dinger equation for a radial wave function in a
central field is \begin{equation}\label{hamH}
\left[\frac{\partial^2}{\partial r^2} +
\frac{1}{r}\frac{\partial}{\partial r} - \frac{m^2}{r^2} -2U(r)+2E
\right]\varphi(r)=0.
\end{equation}
 For the Coulomb potential, $U(r)=-Z/r$, the ground state is
well-known:
\begin{equation}
E_1=-2Z^2,\quad \varphi_{1}(r)=\frac{4Z}{\sqrt{2\pi}}e^{-2rZ}.
\end{equation}
%%%%%%%%%%%%%%%%%%%%%%%%%%%%%%%%%%%%%%%%%%%%%%%%%%%%%%%%%%%%%%%%%%%%
Let us consider now the potential\cite{pon99a}
\begin{equation}\label{Pon}
U(r)=-1/\sqrt{r^2+d^2}.
\end{equation}
The ground state energy and the eigenfunction for Eq.~(\ref{hamH})
with the potential (\ref{Pon}) are presented in  Fig.~\ref{fig3}.
%%%%%%%%%%%%%%%%%%%%%%%%%%%%%%%%%%%%%%%%%%%
%%%%%%%%%%%%%%%%%%%%%%%%%%%%%%%%%%%%%%%%%%%

It turns out that the numerical results are in an extremely good
agreement ($\sim 0.01\%$) with the results of variational method
for the trial function
\begin{equation}\label{varmethtf}
\varphi_{trial}=
\frac{2\exp(d/\lambda)}{\sqrt{\lambda(\lambda+2d)}}
\exp(-\sqrt{r^2+d^2}/\lambda),
\end{equation}
where $\lambda$ is the only variational parameter. The
corresponding value for ground state energy is obtained from
minimization of the functional
\begin{equation}\label{enerexpect}
E_X(\lambda)=-\frac{2}{\lambda}\frac{1}{1+2d/\lambda}+\frac{1}{\lambda^2}
\left[1-\frac{(2d/\lambda)^2E_1(1,2d/\lambda)\exp(2d/\lambda)}{1+2d/\lambda}\right].
\end{equation}
Such a good agreement can be explained by the fact that the trial
function (\ref{varmethtf}) has correct asymptotic behavior in both
limits of small and large $r$.

 Far away from the ``atomic residue''  $U(r)\sim -1/r$,
the  wave function for the $s$-state obeys the Schr\"{o}dinger
equation
$$-\frac{\Delta}{2}\varphi-\frac{1}{r}\varphi=-\frac{\alpha^2}{2}\varphi$$
with the solution
 \be\label{assymH}
\varphi(r)=A_{\alpha}r^{1/\alpha-1/2}e^{-\alpha r},
 \ee
  where terms of the order of $\varphi/r^2$ are neglected. Here
$A_{\alpha}$ and $\alpha$ are atomic parameters. Their magnitudes
are determined by the behavior of an electron inside the ``atom''.

Simple analytical estimates for the cases of small and large $d$
for ground state energy and the asymptotic coefficient
$A_{\alpha}$ can be made.
 When $d\ll 1$ the potential is only slightly different from the Coulomb
potential and the contribution to the energy can be obtained using
perturbation theory: \be\label{pe} E_X=-2+\langle 0|V|0\rangle
\approx -2+16d\left(1+2d\ln(2d)\right). \ee Note, that due to the
big factor in front of $d$ in Eq.~(\ref{pe}), the convergence
radius for the perturbation series is small enough.

 Assuming that the asymptotic form (\ref{assymH})
of the wave function is valid for all values of $r$, we obtain the
following
 simple estimate for $A_{\alpha}$
\be\label{Anorm} 2\pi
A_{\alpha}^2\int_0^{\infty}r^{2/\alpha}\exp(-2\alpha r)\,dr=1
\Rightarrow A_{\alpha}=\frac{2^{1/\alpha-1/2}\alpha^{1/\alpha+1}}
{\sqrt{\pi\Gamma(2/\alpha)}}. \ee
 In the opposite case, when $d\gg 1$ the solution has to be close
to the oscillatory one \be\label{oscen} E_X=
-\frac{1}{d}+\frac{1}{d^{3/2}}(n+1), \ee where $n=0,1,2...$. The
main contribution to the normalization comes from the Gaussian
part of the wave function because the contribution of the Coulomb
tails is negligible. One obtains for the state with $n=0$
\be\label{Aalosc}
\varphi(r)\approx\varphi_{osc}(r)=\sqrt{\frac{\omega}{\pi}}\exp\left(-\omega
r^2/2 \right), \ee
where $\omega=1/d^{3/2}$. Then, \be\label{next}
A_{\alpha}=\frac{1}{\sqrt{\pi}d^{3/4}}. \ee

\section{\label{sec:Ap2}Variational method for excitons in $\delta$-function quantum wells}
Following standard procedures the envelope variational exciton
wave function in a quantum well can be presented as a product of
three terms,
\begin{equation}\label{totalvwf}
\Psi(z_e,z_h,r;\lambda_i)=\chi_e(z_e)\chi_h(z_h)\phi(r,z_e,z_h;\lambda_i),
\end{equation}
where $\lambda_i$ is the set of variational parameters and $\phi$
is the variational wave function which minimizes the total energy
of the Hamiltonian (\ref{H1}). Two other factors
$\chi_{e,h}(z_{e,h})$ are simply  normalized eigenfunctions of the
one-particle electron or hole Hamiltonians of the quantum well:
\begin{eqnarray}
E_{e,h}\chi_{e,h}(z_{e,h})&=&\left[-\frac{1}{2m_{e,h}}\frac{\partial^2}{\partial
z_{e,h}^2}
-\alpha_{e,h}\delta(z_{e,h})\right]\chi_{e,h}(z_{e,h}),\nonumber\\
\chi_{e,h}(z_{e,h})&=&\sqrt{\kappa_{e,h}}\exp(-\kappa_{e,h}|z_{e,h}|),\nonumber\\
\kappa_{e,h} &=& m_{e,h}\alpha_{e,h}, \qquad
E_{e,h}=\kappa_{e,h}^2/2m_{e,h}.
\end{eqnarray}
To obtain more confident results we made calculations with three
different trial functions $\psi$:
\begin{eqnarray}
\psi^{(1)}(r;\lambda)&=& \exp(-r/\lambda),\label{tf01}\\
\psi^{(2)}(r;\lambda,\beta)&=&\exp(-\sqrt{r^2+\beta^2}/\lambda),\label{tf02}\\
\psi^{(3)}(r,|z_e-z_h|;\lambda,\beta)&=&\exp(-\sqrt{r^2+\beta^2(z_e-z_h)^2}/\lambda),\label{tf03}
\end{eqnarray}
The first two functions are independent of $z$ coordinates, while
the third one is non-separable with respect to $z$. The first wave
function has one variational parameter $\lambda$, and two others
have two variational parameters $\lambda$ and $\beta$.

The total exciton energy is the minimum of the functional
\begin{equation}\label{minfunct}
  E=\frac{\langle \Psi|\hat{H}|\Psi\rangle}{\langle
  \Psi|\Psi\rangle}.
\end{equation}
For the first two trial functions, which are independent of $z$,
the functional (\ref{minfunct}) can be further simplified. In this
case, the energy can be presented as
\begin{equation}\label{minfunct01}
  E=E_e+E_h+\overline{K}/N+\overline{V}/N,
\end{equation}
where
\begin{eqnarray}
N&=&\langle\Psi|\Psi\rangle,
\equiv\langle\psi(r)|\psi(r)\rangle\,\label{Normcon}\\
\overline{K}&=&\langle\psi(r)|\hat{K_r}|\psi(r)\rangle\ \equiv
\frac{1}{2}\int \left(\nabla \psi\right)^2 dA
,\label{Kaver}\\
\overline{V}&=&-\langle\Psi|\frac{1}{\sqrt{r^2+(z_e-z_h)^2}}|\Psi\rangle\
\equiv  \int \left(V_{\textrm{eff}}(r) |\psi(r)|^2\right)
dA,\label{Vaver}
\end{eqnarray}
Calculations for the trial function $\psi^{(2)}$ yield
\begin{eqnarray}
N^{(2)}&=&\frac{\exp(-2\beta/\lambda)\lambda(\lambda+2\beta)}{4},\nonumber\\
\overline{K}^{(2)}&=&\frac{\exp(-2\beta/\lambda)}{8}\left[1+\frac{2\beta}{\lambda}
-\left(\frac{2\beta}{\lambda}\right)^2
\textrm{E}_1(2\beta/\lambda)\exp(2\beta/\lambda)
\right],\nonumber\\
\overline{V}^{(2)}&=&-\frac{2\kappa_1\kappa_2}{\kappa_2^2-\kappa_1^2}
\left[\kappa_2V(\kappa_1,\lambda,\beta)- \kappa_1
V(\kappa_2,\lambda,\beta)\right],
\end{eqnarray}
where $\textrm{E}_1(x)$ is the exponential
integral\cite{Abramowitz} and
\begin{equation}\label{Vlambda0}
  V(\kappa,\lambda,\beta)=
  \int_{0}^{\pi/2}\cos(\phi)\;d\phi \int_{0}^{\infty}\exp(-2\kappa R\sin(\phi)-2\sqrt{R^2\cos^2(\phi)+\beta^2}/\lambda)R\;dR.
\end{equation}
For the first trial function $\psi^{(1)}(r;\lambda)\equiv
\psi^{(2)}(r;\lambda,0)$ all integrals have analytical
expressions:
\begin{eqnarray}
N^{(1)}&=&\frac{\lambda^2}{4},\nonumber\\
\overline{K}^{(1)}&=&\frac{1}{8},\nonumber\\
V(\kappa,\lambda,0)&=&\frac{\lambda^2}{4}\frac{1}{1+\lambda^2\kappa^2}\left[
\lambda\kappa -1
+\frac{1}{\sqrt{1+\lambda^2\kappa^2}}\ln\left(\frac{1}{\lambda\kappa}\frac{\sqrt{1+\lambda^2\kappa^2}+1}
{\sqrt{1+\lambda^2\kappa^2}-\lambda\kappa} \right)\right].
\end{eqnarray}
Integrals for the non-separable trial function $\psi^{(3)}$ can be
numerically estimated following the procedure described in
Ref.~\onlinecite{Harrison96a,Harrison99a}

%%%%%%%%%% Produces the bibliography via BibTeX. %%%%%%%%%%%%
%\bibliography{exciton}

\begin{thebibliography}{26}
\expandafter\ifx\csname
natexlab\endcsname\relax\def\natexlab#1{#1}\fi
\expandafter\ifx\csname bibnamefont\endcsname\relax
  \def\bibnamefont#1{#1}\fi
\expandafter\ifx\csname bibfnamefont\endcsname\relax
  \def\bibfnamefont#1{#1}\fi
\expandafter\ifx\csname citenamefont\endcsname\relax
  \def\citenamefont#1{#1}\fi
\expandafter\ifx\csname url\endcsname\relax
  \def\url#1{\texttt{#1}}\fi
\expandafter\ifx\csname
urlprefix\endcsname\relax\def\urlprefix{URL }\fi
\providecommand{\bibinfo}[2]{#2}
\providecommand{\eprint}[2][]{\url{#2}}

\bibitem[{\citenamefont{Harrison}(1999)}]{Harrison99a}
\bibinfo{editor}{\bibfnamefont{P.}~\bibnamefont{Harrison}}, ed.,
  \emph{\bibinfo{title}{Quantum wells, wires, and dots: theoretical and
  computational physics}} (\bibinfo{publisher}{John Wiley and Sons},
  \bibinfo{address}{Chichester}, \bibinfo{year}{1999}).

\bibitem[{\citenamefont{Burstein and Weisbuch}(1995)}]{Burstein}
\bibinfo{editor}{\bibfnamefont{E.}~\bibnamefont{Burstein}} \bibnamefont{and}
  \bibinfo{editor}{\bibfnamefont{C.}~\bibnamefont{Weisbuch}}, eds.,
  \emph{\bibinfo{title}{Confined Electrons and Photons: New Physics and
  Applications}} (\bibinfo{publisher}{Plenum Press}, \bibinfo{address}{New
  York}, \bibinfo{year}{1995}).

\bibitem[{\citenamefont{Mendez and von Klitzing}(1989)}]{Mendez}
\bibinfo{editor}{\bibfnamefont{E.~E.} \bibnamefont{Mendez}} \bibnamefont{and}
  \bibinfo{editor}{\bibfnamefont{K.}~\bibnamefont{von Klitzing}}, eds.,
  \emph{\bibinfo{title}{Physics and Applications of Quantum Wells and
  Superlattices}} (\bibinfo{publisher}{Plenum Press}, \bibinfo{address}{New
  York}, \bibinfo{year}{1989}).

\bibitem[{\citenamefont{Nolte}(1999)}]{Nolte}
\bibinfo{author}{\bibfnamefont{D.~D.} \bibnamefont{Nolte}},
  \bibinfo{journal}{Journal of Applied Physics} \textbf{\bibinfo{volume}{85}},
  \bibinfo{pages}{6259} (\bibinfo{year}{1999}).

\bibitem[{\citenamefont{Miller et~al.}(1981)\citenamefont{Miller, Kleinman,
  Tsang, and Gossard}}]{Miller81}
\bibinfo{author}{\bibfnamefont{R.~C.} \bibnamefont{Miller}},
  \bibinfo{author}{\bibfnamefont{D.~A.} \bibnamefont{Kleinman}},
  \bibinfo{author}{\bibfnamefont{W.~T.} \bibnamefont{Tsang}}, \bibnamefont{and}
  \bibinfo{author}{\bibfnamefont{A.~C.} \bibnamefont{Gossard}},
  \bibinfo{journal}{Phys. Rev. B} \textbf{\bibinfo{volume}{24}},
  \bibinfo{pages}{1134} (\bibinfo{year}{1981}).

\bibitem[{\citenamefont{Bastard et~al.}(1982)\citenamefont{Bastard, Mendez,
  Chang, and Esaki}}]{Bastard82a}
\bibinfo{author}{\bibfnamefont{G.}~\bibnamefont{Bastard}},
  \bibinfo{author}{\bibfnamefont{E.~E.} \bibnamefont{Mendez}},
  \bibinfo{author}{\bibfnamefont{L.~L.} \bibnamefont{Chang}}, \bibnamefont{and}
  \bibinfo{author}{\bibfnamefont{L.}~\bibnamefont{Esaki}},
  \bibinfo{journal}{Phys. Rev. B} \textbf{\bibinfo{volume}{26}},
  \bibinfo{pages}{1974} (\bibinfo{year}{1982}).

\bibitem[{\citenamefont{Greene et~al.}(1984)\citenamefont{Greene, Bajaj, and
  Phelps}}]{Green84}
\bibinfo{author}{\bibfnamefont{R.~L.} \bibnamefont{Greene}},
  \bibinfo{author}{\bibfnamefont{K.~K.} \bibnamefont{Bajaj}}, \bibnamefont{and}
  \bibinfo{author}{\bibfnamefont{D.~E.} \bibnamefont{Phelps}},
  \bibinfo{journal}{Phys. Rev. B} \textbf{\bibinfo{volume}{29}},
  \bibinfo{pages}{1807} (\bibinfo{year}{1984}).

\bibitem[{\citenamefont{Efros}(1986)}]{Efros86a}
\bibinfo{author}{\bibfnamefont{A.~L.} \bibnamefont{Efros}},
  \bibinfo{journal}{Sov. Phys. Semicond.} \textbf{\bibinfo{volume}{20}},
  \bibinfo{pages}{808} (\bibinfo{year}{1986}).

\bibitem[{\citenamefont{Andreani and Pasquarello}(1990)}]{Andreani90a}
\bibinfo{author}{\bibfnamefont{L.~C.} \bibnamefont{Andreani}} \bibnamefont{and}
  \bibinfo{author}{\bibfnamefont{A.}~\bibnamefont{Pasquarello}},
  \bibinfo{journal}{Phys. Rev. B} \textbf{\bibinfo{volume}{42}},
  \bibinfo{pages}{8928} (\bibinfo{year}{1990}).

\bibitem[{\citenamefont{Gerlach et~al.}(1998)\citenamefont{Gerlach, Wusthoff,
  Dzero, and Smondyrev}}]{Gerlach98a}
\bibinfo{author}{\bibfnamefont{B.}~\bibnamefont{Gerlach}},
  \bibinfo{author}{\bibfnamefont{J.}~\bibnamefont{Wusthoff}},
  \bibinfo{author}{\bibfnamefont{M.~O.} \bibnamefont{Dzero}}, \bibnamefont{and}
  \bibinfo{author}{\bibfnamefont{M.~A.} \bibnamefont{Smondyrev}},
  \bibinfo{journal}{Phys. Rev. B} \textbf{\bibinfo{volume}{58}},
  \bibinfo{pages}{10568} (\bibinfo{year}{1998}).

\bibitem[{\citenamefont{Iotti and Andreani}(1997)}]{Andreani97a}
\bibinfo{author}{\bibfnamefont{R.~C.} \bibnamefont{Iotti}} \bibnamefont{and}
  \bibinfo{author}{\bibfnamefont{L.~C.} \bibnamefont{Andreani}},
  \bibinfo{journal}{Phys. Rev. B} \textbf{\bibinfo{volume}{56}},
  \bibinfo{pages}{3922} (\bibinfo{year}{1997}).

\bibitem[{\citenamefont{Kossut et~al.}(1997)\citenamefont{Kossut, Furdyna, and
  Dobrowolska}}]{Kossut97a}
\bibinfo{author}{\bibfnamefont{J.}~\bibnamefont{Kossut}},
  \bibinfo{author}{\bibfnamefont{J.~K.} \bibnamefont{Furdyna}},
  \bibnamefont{and}
  \bibinfo{author}{\bibfnamefont{M.}~\bibnamefont{Dobrowolska}},
  \bibinfo{journal}{Phys. Rev. B} \textbf{\bibinfo{volume}{56}},
  \bibinfo{pages}{9775} (\bibinfo{year}{1997}).

\bibitem[{\citenamefont{Harrison et~al.}(1996)\citenamefont{Harrison, Piorek,
  Hagston, and Stirner}}]{Harrison96a}
\bibinfo{author}{\bibfnamefont{P.}~\bibnamefont{Harrison}},
  \bibinfo{author}{\bibfnamefont{T.}~\bibnamefont{Piorek}},
  \bibinfo{author}{\bibfnamefont{W.~E.} \bibnamefont{Hagston}},
  \bibnamefont{and} \bibinfo{author}{\bibfnamefont{T.}~\bibnamefont{Stirner}},
  \bibinfo{journal}{Superlatt. Microstruct.} \textbf{\bibinfo{volume}{20}},
  \bibinfo{pages}{45} (\bibinfo{year}{1996}).

\bibitem[{\citenamefont{de~Leon and Laikhtman}(2000)}]{Laikhtman00a}
\bibinfo{author}{\bibfnamefont{S.}~\bibnamefont{de~Leon}} \bibnamefont{and}
  \bibinfo{author}{\bibfnamefont{B.}~\bibnamefont{Laikhtman}},
  \bibinfo{journal}{Phys. Rev. B} \textbf{\bibinfo{volume}{61}},
  \bibinfo{pages}{2874} (\bibinfo{year}{2000}).

\bibitem[{\citenamefont{Ekenberg and Altarelli}(1987)}]{Ekenberg87}
\bibinfo{author}{\bibfnamefont{U.}~\bibnamefont{Ekenberg}} \bibnamefont{and}
  \bibinfo{author}{\bibfnamefont{M.}~\bibnamefont{Altarelli}},
  \bibinfo{journal}{Phys. Rev. B} \textbf{\bibinfo{volume}{35}},
  \bibinfo{pages}{7585} (\bibinfo{year}{1987}).

\bibitem[{\citenamefont{Chang et~al.}(1988)\citenamefont{Chang, Nurmikko, Wu,
  Kolodziejski, and Gunshor}}]{Chang88}
\bibinfo{author}{\bibfnamefont{S.~K.} \bibnamefont{Chang}},
  \bibinfo{author}{\bibfnamefont{A.~V.} \bibnamefont{Nurmikko}},
  \bibinfo{author}{\bibfnamefont{J.~W.} \bibnamefont{Wu}},
  \bibinfo{author}{\bibfnamefont{L.~A.} \bibnamefont{Kolodziejski}},
  \bibnamefont{and} \bibinfo{author}{\bibfnamefont{R.~L.}
  \bibnamefont{Gunshor}}, \bibinfo{journal}{Phys. Rev. B}
  \textbf{\bibinfo{volume}{37}}, \bibinfo{pages}{1191} (\bibinfo{year}{1988}).

\bibitem[{\citenamefont{Warnock et~al.}(1993)\citenamefont{Warnock, Jonker,
  Petrou, Chou, and Liu}}]{Warnock93a}
\bibinfo{author}{\bibfnamefont{J.}~\bibnamefont{Warnock}},
  \bibinfo{author}{\bibfnamefont{B.~T.} \bibnamefont{Jonker}},
  \bibinfo{author}{\bibfnamefont{A.}~\bibnamefont{Petrou}},
  \bibinfo{author}{\bibfnamefont{W.~C.} \bibnamefont{Chou}}, \bibnamefont{and}
  \bibinfo{author}{\bibfnamefont{X.}~\bibnamefont{Liu}},
  \bibinfo{journal}{Phys. Rev. B} \textbf{\bibinfo{volume}{48}},
  \bibinfo{pages}{17321} (\bibinfo{year}{1993}).

\bibitem[{\citenamefont{Piorek et~al.}(1995)\citenamefont{Piorek, Harrison, ,
  and Hagston}}]{Harrison95a}
\bibinfo{author}{\bibfnamefont{T.}~\bibnamefont{Piorek}},
  \bibinfo{author}{\bibfnamefont{P.}~\bibnamefont{Harrison}}, ,
  \bibnamefont{and} \bibinfo{author}{\bibfnamefont{W.~E.}
  \bibnamefont{Hagston}}, \bibinfo{journal}{Phys. Rev. B}
  \textbf{\bibinfo{volume}{52}}, \bibinfo{pages}{14111} (\bibinfo{year}{1995}).

\bibitem[{\citenamefont{Voliotis et~al.}(1995)\citenamefont{Voliotis, Grousson,
  Lavallard, and Planel}}]{Voliotis95a}
\bibinfo{author}{\bibfnamefont{V.}~\bibnamefont{Voliotis}},
  \bibinfo{author}{\bibfnamefont{R.}~\bibnamefont{Grousson}},
  \bibinfo{author}{\bibfnamefont{P.}~\bibnamefont{Lavallard}},
  \bibnamefont{and} \bibinfo{author}{\bibfnamefont{R.}~\bibnamefont{Planel}},
  \bibinfo{journal}{Phys. Rev. B} \textbf{\bibinfo{volume}{52}},
  \bibinfo{pages}{10725} (\bibinfo{year}{1995}).

\bibitem[{\citenamefont{Stahl and Balslev}(1987)}]{Stahl87}
\bibinfo{author}{\bibfnamefont{A.}~\bibnamefont{Stahl}} \bibnamefont{and}
  \bibinfo{author}{\bibfnamefont{I.}~\bibnamefont{Balslev}},
  \emph{\bibinfo{title}{Electrodynamics of the Semiconductor Band Edge}}
  (\bibinfo{publisher}{Springer}, \bibinfo{address}{Berlin},
  \bibinfo{year}{1987}).

\bibitem[{\citenamefont{Balslev et~al.}(1989)\citenamefont{Balslev, Zimmermann,
  and Stahl}}]{Balslev89a}
\bibinfo{author}{\bibfnamefont{I.}~\bibnamefont{Balslev}},
  \bibinfo{author}{\bibfnamefont{R.}~\bibnamefont{Zimmermann}},
  \bibnamefont{and} \bibinfo{author}{\bibfnamefont{A.}~\bibnamefont{Stahl}},
  \bibinfo{journal}{Phys. Rev. B} \textbf{\bibinfo{volume}{40}},
  \bibinfo{pages}{4095} (\bibinfo{year}{1989}).

\bibitem[{\citenamefont{Merbach et~al.}(1998)\citenamefont{Merbach, Sch\"{o}ll,
  Ebeling, Michler, and Gutowski}}]{Merbach98}
\bibinfo{author}{\bibfnamefont{D.}~\bibnamefont{Merbach}},
  \bibinfo{author}{\bibfnamefont{E.}~\bibnamefont{Sch\"{o}ll}},
  \bibinfo{author}{\bibfnamefont{W.}~\bibnamefont{Ebeling}},
  \bibinfo{author}{\bibfnamefont{P.}~\bibnamefont{Michler}}, \bibnamefont{and}
  \bibinfo{author}{\bibfnamefont{J.}~\bibnamefont{Gutowski}},
  \bibinfo{journal}{Phys. Rev. B} \textbf{\bibinfo{volume}{58}},
  \bibinfo{pages}{10709} (\bibinfo{year}{1998}).

\bibitem[{\citenamefont{Castella and Wilkins}(1998)}]{Castella98}
\bibinfo{author}{\bibfnamefont{H.}~\bibnamefont{Castella}} \bibnamefont{and}
  \bibinfo{author}{\bibfnamefont{J.~W.} \bibnamefont{Wilkins}},
  \bibinfo{journal}{Phys. Rev. B} \textbf{\bibinfo{volume}{58}},
  \bibinfo{pages}{16186} (\bibinfo{year}{1998}).

\bibitem[{\citenamefont{Miller et~al.}(1985)\citenamefont{Miller, Chemla,
  Damen, Gossard, Wiegmann, Wood, and Burrus}}]{Miller85a}
\bibinfo{author}{\bibfnamefont{D.~A.~B.} \bibnamefont{Miller}},
  \bibinfo{author}{\bibfnamefont{D.~S.} \bibnamefont{Chemla}},
  \bibinfo{author}{\bibfnamefont{T.~C.} \bibnamefont{Damen}},
  \bibinfo{author}{\bibfnamefont{A.~C.} \bibnamefont{Gossard}},
  \bibinfo{author}{\bibfnamefont{W.}~\bibnamefont{Wiegmann}},
  \bibinfo{author}{\bibfnamefont{T.~H.} \bibnamefont{Wood}}, \bibnamefont{and}
  \bibinfo{author}{\bibfnamefont{C.~A.} \bibnamefont{Burrus}},
  \bibinfo{journal}{Phys. Rev. B} \textbf{\bibinfo{volume}{32}},
  \bibinfo{pages}{1043} (\bibinfo{year}{1985}).

\bibitem[{\citenamefont{Abramowitz and Stegun}(1965)}]{Abramowitz}
\bibinfo{editor}{\bibfnamefont{M.}~\bibnamefont{Abramowitz}} \bibnamefont{and}
  \bibinfo{editor}{\bibfnamefont{I.~A.} \bibnamefont{Stegun}}, eds.,
  \emph{\bibinfo{title}{Handbook of Mathematical Functions}}
  (\bibinfo{publisher}{Dover}, \bibinfo{address}{New York},
  \bibinfo{year}{1965}).

\bibitem[{\citenamefont{Ponomarev et~al.}(1999)\citenamefont{Ponomarev,
  Flambaum, and Efros}}]{pon99a}
\bibinfo{author}{\bibfnamefont{I.~V.} \bibnamefont{Ponomarev}},
  \bibinfo{author}{\bibfnamefont{V.~V.} \bibnamefont{Flambaum}},
  \bibnamefont{and} \bibinfo{author}{\bibfnamefont{A.~L.} \bibnamefont{Efros}},
  \bibinfo{journal}{Phys. Rev. B} \textbf{\bibinfo{volume}{60}},
  \bibinfo{pages}{5485} (\bibinfo{year}{1999}).

\end{thebibliography}

\end{document}